\documentclass[amsmath,amssymb,prl,superscriptaddress,reprint,showpacs,longbibliography]{revtex4-1}
\usepackage{graphicx}
\usepackage{dcolumn}
\usepackage[colorlinks=true,linkcolor=blue,citecolor=blue,urlcolor=blue]{hyperref}
\usepackage[usenames,dvipsnames]{xcolor}
\usepackage{soul}
\usepackage{braket}
\usepackage{bm}
\usepackage{upgreek}
\usepackage{mathtools}
\usepackage{bbm}
\usepackage{multirow}

\renewcommand{\vec}[1]{{\bf #1}}

\newcommand{\gaurav}[1]{\textcolor{blue}{[GC: #1]}}
\newcommand{\ivar}[1]{\textcolor{red}{IM: #1}}

\begin{document}

\title{Superconductivity from domain wall fluctuations in sliding ferroelectrics}


\author{Gaurav Chaudhary}
\email{gc674@cam.ac.uk}
\affiliation{TCM Group, Cavendish Laboratory, University of Cambridge, J. J. Thomson Avenue, Cambridge CB3 0HE, United Kingdom}

\author{Ivar Martin}
\email{ivar@anl.gov }
\affiliation{Materials Science Division, Argonne National Laboratory, Lemont, IL 60439, USA}

\date{\today}



\begin{abstract} 
Bilayers of two-dimensional van der Waals materials that lack an inversion centre can show a novel form of ferroelectricity, where certain stacking arrangements of the two layers lead to an interlayer polarization. 
Under an external out-of-plane electric field, a relative sliding between the two layers can occur accompanied by an inter-layer charge transfer and a ferroelectric switching. 
We show that the domain walls that mediate ferroelectric switching are a locus of strong attractive interactions between electrons. 
The attraction is mediated by the ferroelectric domain wall fluctuations, effectively driven by the soft interlayer shear phonon.
We comment on the possible relevance of this attraction mechanism to the recent observation of an interplay between sliding ferroelectricity and superconductivity in bilayer $\text{T$_d$-MoTe}_2$. We also discuss the possible role of this mechanism in the superconductivity of moir\'e bilayers.  
\end{abstract}

\maketitle

\textit{Introduction}- 
Recently ferroelectricity has been discovered in several stacking-engineered bilayers and twisted bilayers of van der Waals materials\cite{Fei2018, Yasuda2021, Stern2021, Barrera2021, Wang2022}. In bilayers of non-elemental materials that lack an inversion symmetry at the monolayer level, certain bilayer stacking arrangements lead to interlayer charge transfer. Even when the layers themselves are metallic, such charge imbalance can be interpreted as ferroelectricity. 
For example, AB stacked bilayers where A and B are different elements may lead to an interlayer charge transfer because of the different electron affinity of A and B elements. 
Remarkably, the polarization direction can be switched via the application of a transverse electric field, which reverses the stacking order to BA; this is the mechanism of the \textit{sliding ferroelectricity}. 
Similarly, moir\'e engineering of these materials via small relative twist or strain leads to moir\'e polar domains~\cite{Enaldiev2021,Enaldiev2022,Bennett2022,Bennett2023}. 
Apart from bilayers of non-elemental materials, ferroelectricity was also discovered in graphene bilayers, possibly due to strong electron-electron interactions~\cite{Zheng2020, Niu2022, Zheng2023}. 

\begin{figure}[h]
  \includegraphics[width=0.45\textwidth]{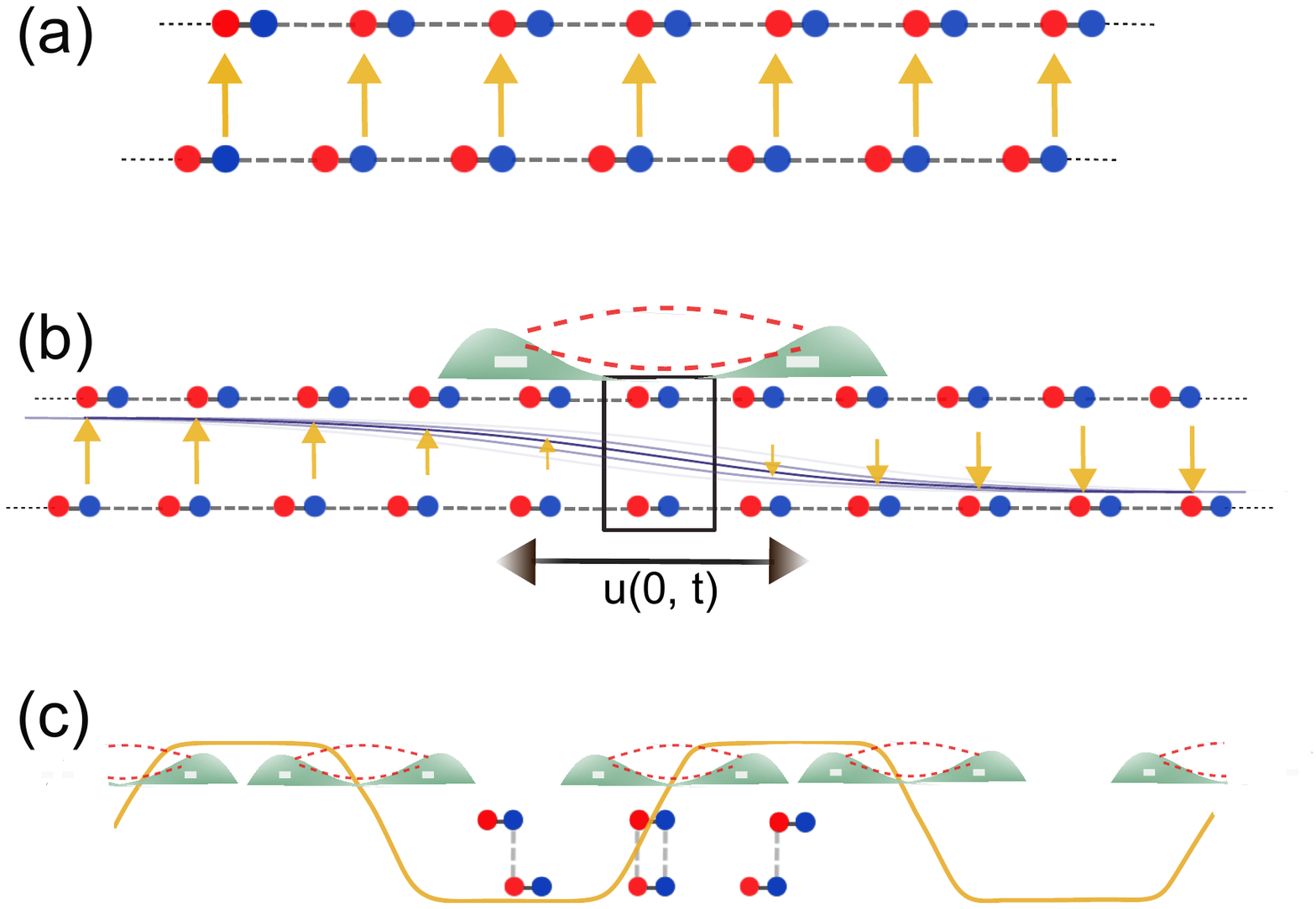}
  \caption{\label{Fig.:1d_schematic}
  Illustration of sliding ferroelectricity in two stacked diatomic chains: (a) When the stacking is AB (or BA), inter-chain(layer) polarization can develop by charge transfer due to different on-site energies of $A$ and $B$ sublattice. 
  (b) When the stacking arrangement changes from AB to BA at the origin, the interlayer polarization points in opposite directions on either side. The fluctuations in the region where polarization vanishes generate effective attractive interactions between electrons, leading to Cooper pair formation as shown in green. (c) A small mismatch between the lattice constant of two chains can create a moir\'e superlattice where interlayer polarization takes moir\'e periodic profile.}
\end{figure}

A unique feature of these two-dimensional ferroelectrics is that they are compatible with the in-plane metallicity and superconductivity. 
In bulk metals or superconductors, a long-range ferroelectric order cannot develop because any such order is screened by mobile charge carriers. 
However, in the two-dimensional ferroelectrics, interlayer polarization can survive the presence of in-plane conduction. Indeed, a ferroelectric metal was discovered in bilayer orthorhombic T$_d$-$
\text{WTe}_2$~\cite{Fei2018, Barrera2021}. 
Even more surprisingly an intriguing electric-field switching of superconductivity was recently observed in T$_d$-$\text{MoTe}_2$~\cite{Jindal2023}. 
Under an applied external out-of-plane electric field, as the bilayers approach the polarization reversal, superconducting $T_c$ is strongly enhanced, followed by a rapid drop as the system becomes fully polarized again. This suggests that superconductivity is enhanced on the domain walls that mediate ferroelectric switching.
Such electric field (gating) control of superconductivity can have groundbreaking applications in superconducting devices.

Superconductivity typically emerges due to an attractive pairing interaction between electrons, mediated by 
an exchange of a soft bosonic mode. 
In the conventional BCS theory of superconductivity, these intermediary bosonic modes are phonons~\cite{Bardeen1957}. In high-temperature cuprate superconductors, it is believed that soft antiferromagnetic fluctuations are the relevant bosonic modes~\cite{Moriya1990}. 
Here, we show that in the metallic sliding ferroelectrics, fluctuations of domain walls separating domains of the opposite polar orders naturally lead to strong effective attractive electron-electron interactions, which favor intra-layer pairing. 
If the domain wall fluctuations are associated with structural fluctuations, i.e. phonons, we can recast this mechanism in terms of an effective electron-phonon coupling. The nature of this coupling is \textit{transverse} piezoelectric, with the induced {tranverse} polarization in the center of the domain wall being proportional to the shear strain between the layers. Dynamically such strain fluctuations can be mediated by the interlayer shear phonons (deriving from the { transverse} bulk phonons with the propagation direction normal to the layers). That is in contrast to the conventional BCS mechanism, which relies on the attraction mediated by the longitudinal optical phonons~\cite{mahan2000many}.
Near the polarization switching transition, the local superconductivity at the domain walls can percolate through the entire system leading to vanishing resistivity, possibly explaining the experimental observations in Ref.~\cite{Jindal2023}.

\textit{Hamiltonian and effective attractive interactions}- To illustrate the mechanism in the simplest possible setting, we start with the two diatomic chains as shown in Fig.~\ref{Fig.:1d_schematic} and described by the Hamiltonian
\begin{align}\label{Eq:H_start}
H = \int dx [\hat{\psi}^{\dagger}(x) \{ H_e (x, \partial_x )  + H_{e-p} (x) \} \hat{\psi}(x) + H_p(x)].
\end{align}
Here $H_e$ accounts for the electron kinetic energy contributions including the interlayer hopping, $H_p (x) = P^2(x)/(2\epsilon) $ is the electrostatic energy for polarization $P(x)$ and permittivity $\epsilon$. Electrons moving in the polarized background experience potential energy $H_{e-p}(x) = D(x) \tau_3 $, where $D(x) = edP(x)/\epsilon $ and $d$ is the interlayer distance.  
The electron operator $\hat{\psi}(x)$ has spin, layer, sublattice pseudospins structure, and $\tau$ Pauli matrices act in the layer basis. 
Insulating systems can also develop in-plane (along the chains) polarization components~\cite{Bennett2023}. 
In a two-dimensional metal, static in-plane polarization vanishes due to screening by the in-plane itinerant charges. 
Therefore, the metallic systems of interest only have interlayer polarization. 

In the following discussion, we focus on the electron-polarization interaction term. 
Since the polarization changes as the system is  deformed from its equilibrium  configuration, we consider a real polarization that has both an explicit spatial dependence and an implicit spatial dependence via a deformation field $u(x)$ that can drive fluctuations in polarisation such that 
\begin{align}\label{Eq:Polar_fluc}
    P(x, u(x)) = P(x, 0) +  \left.\frac{\partial P(x, u)}{\partial u}\right|_{u = 0}   d u(x) +  O(u^2),
\end{align}
where $P(x, 0)$ is the static polarization before the deformation. 
We decompose the Hamiltonian in Eq.~\ref{Eq:H_start} as
\begin{align}\label{Eq:Ham_stat_dy}
    & H = \int d x [ \hat{\psi}^{\dagger}(x) \{H_{0}(x,\partial_x) + H_{1}(x) \}  \hat{\psi}(x) + H_p(x)],
\end{align}
where $H_{0} = H_e + H_{e-p}$ takes into account the electron Hamiltonian and the coupling of static polarization to the electrons. The dynamic deformation enters the Hamiltonian through $H_1 = H_u + H_{e-u}$, where $H_u$ is the Hamiltonian associated with the dynamic deformation itself and $H_{e-u}$ accounts for its coupling to electrons.

As shown in Fig.~\ref{Fig.:1d_schematic} (b), far on either side of the domain wall, where polarization is saturated, the first-order spatial derivative of polarization vanishes such that a local deformation will not induce a change in the polarization to the first order in $u$.  
Whereas, at the domain wall [centred at the origin in Fig.~\ref{Fig.:1d_schematic} (b)], deformation $u(0)$ leads to rapid change in polarization. 
Therefore, we can expect that the strongest coupling between the dynamical deformation and electrons will be achieved in the spatial region where the interlayer polarization switches its direction [See Fig.~\ref{Fig.:1d_schematic} (b)], or in moir\'e bilayers, at the domain walls 
separating regularly arranged regions with opposite polarizations [See Fig.~\ref{Fig.:1d_schematic} (c)].

To consider fluctuations in the deformation field, we take the harmonic approximation and follow the standard procedure to quantize the fluctuations by promoting them to bosonic operators $\hat{a}$ and using the substitution $\hat{u}  = \sqrt{\hbar\omega/(8 \kappa )}(\hat{a} + \hat{a}^{\dagger})$, where $\kappa$ is an effective spring constant of the restoring force and $\omega$ is the fundamental frequency. 
Here, we have assumed that $\hat{a}$-boson is a local vibration mode at the domain wall; for the moment, we assume that they are independent at different domain walls.
The electron-boson Hamiltonian becomes 
\begin{align}\label{Eq:Ham_eu}
    H_{e-u} = \int dx g(x) (\hat{a} + \hat{a}^{\dagger}) \hat{\psi}^{\dagger}(x) \tau_3 \hat{\psi}(x), 
\end{align}
where $g(x) = ed \sqrt{\hbar \omega/ (8 \epsilon^2 \kappa )} \partial P(x, u)/\partial u$. 
We integrate out the $\hat{a}$-boson and obtain an effective electron-electron interaction 
\begin{align}\label{Eq:Ham_ee}
    & H_{e-e} = -\sum_{s,s'}\int dx dx' g(x) g(x')   \frac{2\hbar \omega}{\hbar^2\omega^2- E^2}  \notag\\
    & \hspace{1.5cm}\times \tau_{3,ss}\tau_{3,s's'} \hat{\psi}^{\dagger}_{s}(x)\hat{\psi}^{\dagger}_{s'}(x')\hat{\psi}_{s'}(x')\hat{\psi}_s(x),
\end{align}
where  $E$, and is the electron energy. Clearly, attractive interactions are generated in the regime $E < \hbar \omega$ for the intra-layer Cooper pair channel. 
We also note that the interlayer interactions are repulsive with the same strength. This is a distinctive feature of interaction mediated by the interlayer polarization fluctuations.  
If the electronic states near the domain walls are approximately the layer eigenstates, we can restrict to the intra-layer attractive channel, which is fully decoupled from the interlayer repulsive channel.

\textit{Phonon-induced domain wall fluctuation}- 
To determine whether these attractive interactions can lead to substantial superconductivity in a real system, we require additional input about the microscopic origin of the domain wall fluctuations. We now consider a phonon-based origins of the fluctuations.  
\begin{figure}[t]
  \includegraphics[width=0.45\textwidth]{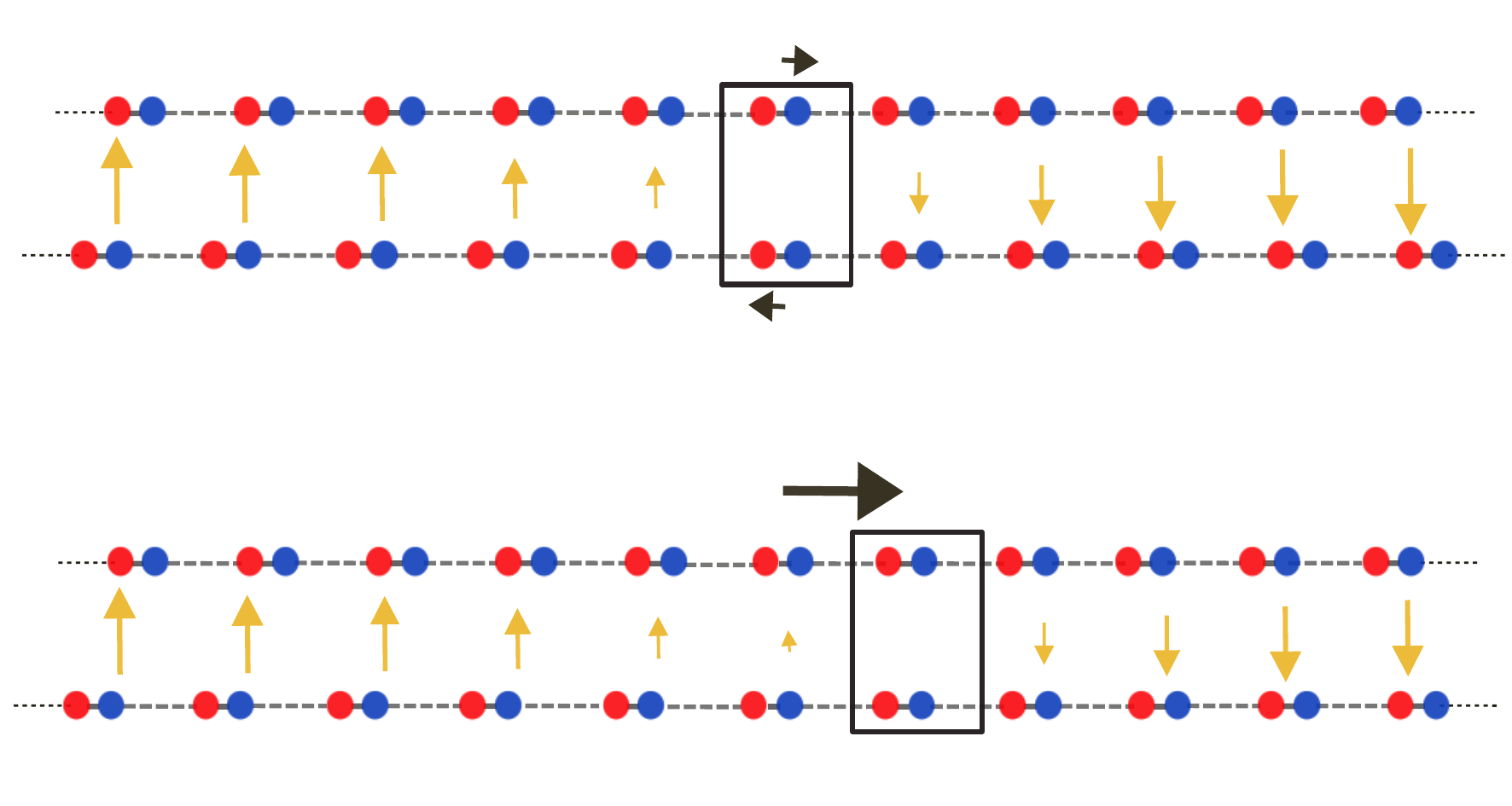}
  \caption{\label{Fig.:1d_slide}
  Illustration of domain wall fluctuations due to interlayer shear phonon. A small displacement of the two layers in opposite directions leads to an enhanced displacement of the domain wall center.}
\end{figure}

Consider a displacement field $u$ and decompose it over in-phase and out-of-phase interlayer displacements, such that $u_{in} = (u_t + u_b)/2 $ and $u_{out} = (u_t-u_b)/2$. Assuming these displacements are independent and acting locally near a domain wall, the resultant displacements of the center of the domain wall are respectively, $u_{dw} = u_{in}$ and $u_{dw} = u_{out} w/a$, where $w$ is the domain wall width and $a$ is the microscopic lattice constant (this follows from the observation that a relative layer shift by a lattice constant leads to a domain wall center shifting by the distance $\sim w$). As a result, the out-of-phase motion of atoms between the layers has a much larger effect on the domain wall position than the in-phase one, leading to significantly stronger coupling between electrons and the fluctuations of $u_{out}$ compared to $u_{in}$ [See Fig.~\ref{Fig.:1d_slide}].
Moreover, the sliding motion mechanism of the ferroelectric transition is indeed driven by the shear phonon. 
Therefore, we now consider the interlayer shear phonon, which is associated with $q=0$ out-of-phase motion of two layers, $ u_{ph} = (u_t - u_b)/2$ [Fig.~\ref{Fig.:1d_slide} (a)]. From the preceding discussion, the domain wall displacement width $u_{dw} = w u_{ph}/a$, leads to a stronger coupling of the shear phonon to electronic states in the domain wall region.
Therefore, in what follows, we will focus on the interlayer shear mode for domain wall fluctuation-induced superconductivity. 
\begin{figure*}[t]
  \includegraphics[width=0.9\textwidth]{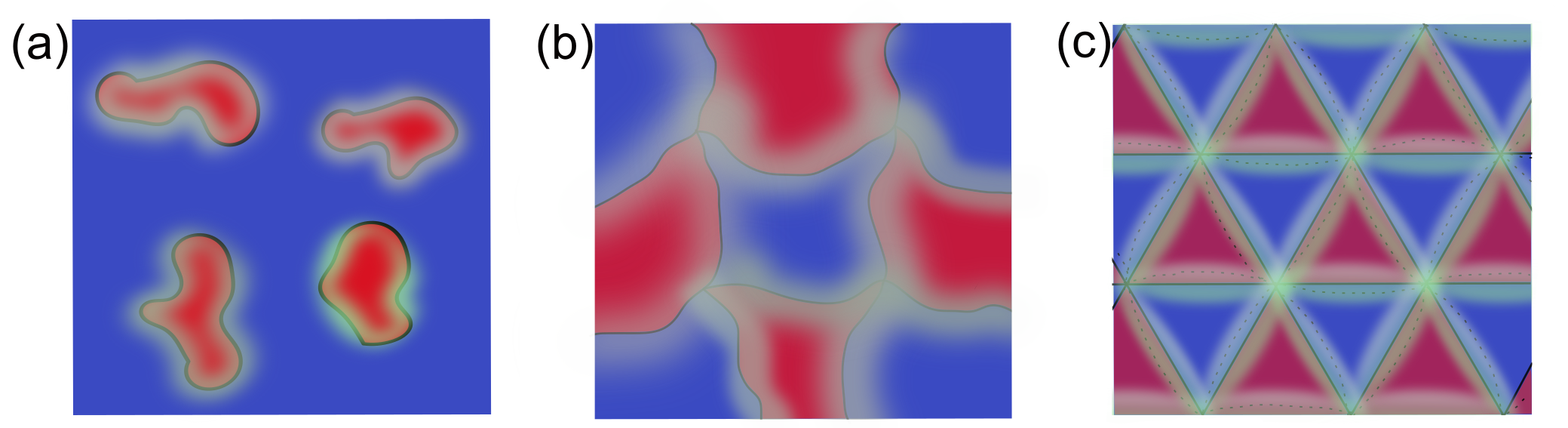}
  \caption{\label{Fig.:2d_schematic}
  Illustration of superconductivity in sliding ferroelectric via pair formation at the domain walls. (a) When the bilayer is polarized, most of it is in $AB (or BA)$ stacking, with smaller disconnected domains of opposite stacking; local pairs form at the domain walls. (b) Near the polarization reversal,  there are equal fractions of domains of both polarizations, with extended domain walls providing a random percolating path for  superconducting transport.  (c) In moire twisted bilayers, the domain walls form an ordered spanning network that supports superconductivity.}
\end{figure*}

Finally, we note that the domain wall fluctuations need not be caused by phonons. 
For example, polarization order parameter fluctuations can occur due to order parameter dynamics in the standard Ginzburg-Landau theory. 
In particular, the interlayer ferroelectric order in bilayer graphene is believed to be primarily due to electronic correlations, rather than the hetero-atomic structure described above, and therefore need not involve a sliding motion during polarization switching~\cite{Zheng2020, Niu2022, Zheng2023}. 
Nevertheless, at the polarization switching transition, the domain wall fluctuations can still lead to local attractive interactions of purely electronic origin. 

\textit{Infinitely-wide domain wall limit.}
Domain walls, by definition, separate energetically stable regions -- in our case, regions with uniform saturated interlayer polarization. They represent saddle-point configurations that would be unstable under conditions of translational invariance. Being unstable, they can offer a unique electronic environment favoring superconductivity. Indeed, as we showed in the previous section,  the coupling constant $dP/du_{out}$ between electrons and the shear phonons is maximized at the center of the domain wall.
Yet, the finite width of the domain wall may lead to suppression of superconductivity, particularly when the width becomes smaller than the superconducting coherence length \cite{PhysRevB.72.060502}. Therefore, the estimate of the upper bound on the pairing interaction and thus $T_c$ can be obtained assuming that the domain wall is infinitely wide, making the system translationally invariant.
This limit allows us to do a direct comparison between the standard polar LO  mechanism \cite{Bardeen1957, frohlich1954electrons} and the shear phonon mechanisms. In this uniform limit, we can recast the effective electron-electron interaction resulting from the shear phonon coupling in the more familiar-looking form
\begin{align}
    & H_{e-e} = -\sum_{\eta,\eta'= \pm} \int dk dq \frac{g^2(q)}{\hbar\omega(q)} \hat{\phi}^{\dagger}_{-\eta} (k-q)\hat{\phi}^{\dagger}_{-\eta'} (-k+q)\notag\\
    &\hspace{4cm} \times \hat{\phi}_{\eta'}(-k)\hat{\phi}_{\eta}(k).
\end{align}
Here, $\hat{\phi_{\pm}} = (\psi_s \pm \psi_{\bar s})/\sqrt2$ are the electron operators for inter-layer bonding/anti-bonding eigenstates, which are approximately the band eigenstates in this limit;  the coupling constant is $g(q) = 2P_0ed \sqrt{\hbar/(2m  \omega(q)a^2 \epsilon^2)}$, where $P_0$ is the saturation value of the interlayer polarization. Notice that in this scenario not only are the interactions uniformly present throughout the system, suggesting phase coherence, but they are also fully attractive at the Fermi level. 
Notably, the effective coupling scales quadratically with the saturation value of interlayer polarization, which is achieved when the layers are AB (BA) stacked.

\textit{Estimates of electron-phonon coupling strength.} Using the approximation of an infinitely wide domain wall we can make an estimate of the strength of the proposed mechanism.
Even though free carriers screen a static polarization, fluctuating polarization is precisely what generates Fr\"olich-like electron-phonon coupling term~\cite{frohlich1954electrons}. 
In the sliding ferroelectrics, dynamical polarization can be directed either in-plane or, when induced by the interlayer charge transfer between layers, out-of-plane. 
The ratio of these polarizations, assuming that they are driven by very similar phonon modes, can be estimated as  $P_z/P_x\sim d/a $. 
Generally, in the van der Waals bilayers $d> a$. 
For example, in $\text{WTe}_2$, the $c = 15.4\AA$, compared to $a= 3.5\AA$ and $b=6.34\AA$. 
Therefore, the ratio of the corresponding electron-electron attraction strengths can be estimated as  $\sim (d/a)^2$.
It may in fact be even larger since the in-plane polarization fluctuations are dynamically screened by the itinerant electrons.

In addition to modifying interlayer polarization, the shear phonons also affect hybridization between the vdW layers. It is instructive to compare the relative strengths of these two types of electron-phonon coupling.
Changing the stacking of layers can change interlayer hybridization energy by approximately $100~\text{meV}$ for the typical vdW systems. This is therefore approximately the difference in energy per electron located within a stable stacking domain or at the domain wall, which corresponds to a relative layer displacement by approximately a lattice constant. 
To compare with the corresponding electrostatic energies in the  polar metallic phase, we can take the example of in bilayer $T_d$-$\text{WTe}_2$, where experiments and first-principles calculations estimate $P\sim P_0
\times 10^{-4}  C \cdot m^{-2} $, where $P_0
\sim 2-6$~\cite{Wu2021,Fei2018,Barrera2021}.
Simple estimates suggest that it can lead to a typical coupling energy per electron is $V \sim 100 c\, \text{meV}$. Here $c=(P_0 d/\epsilon_r)^2$ is an  $O(1)$ proportionality constant for $d$ measured in $nm$ and $\epsilon_r$ is the dielectric constant. 
Therefore we find that in the domain wall region, these two coupling mechanisms to the shear phonon have comparable strengths. 
In the systems that develop interlayer polarization, the mere fact of the presence of polarization implies that the polarization energy dominates the hybridization (since hybridization favors equal charge distribution between layers). 
In real systems, both effects can work in conjunction to provide an increased electron-to-shear phonon coupling at the stacking domain walls in vdW bilayers. 
We note in passing, that in non-polar 
vdW materials, such as multilayer graphene-based superconductors, the just-described modulation of the interlayer hybridization in the domain wall regions by the shear phonon can be a viable mechanism of local pairing.

In addition to the strength of the electron-phonon coupling discussed above, superconducting transition temperature depends on several other ingredients, which we discuss now.
In the weak-to-intermediate coupling superconductors, the critical temperature follows $T_c \sim \hbar\omega_D \exp -(1+\lambda)/(\lambda-\mu^*)$, where $\omega_D$ is the typical frequency of the most pairing-relevant phonon, $\lambda$ is a dimensionless coupling constant, and $\mu^*$ is the Coulomb pseudopotential~\cite{McMillan1968}. 
Because the $T_c$ is very sensitive to the coupling constants, we only discuss here the distinctive qualitative features of the domain wall mechanism. 
If we neglect the phonon dispersion, the coupling constant can be decomposed as $\lambda = 4\mathcal{N}_F V/(m\omega_D^2) $\footnote{The $q=0$ frequency of the shear phonon is determined by the weak interlayer spring constant. However, its dispersion is predominantly governed by the stiff intralayer spring constant. Therefore, these phonons are highly dispersive and neglecting their dispersion is not a very good approximation.}, where $\mathcal{N}_F$ is the electron density of state at the Fermi level and $V ~\sim 4 (edP_0/a \epsilon)^2 $ quantifies the polarization energy change across the domain wall, $m$ is the unit cell mass, and $\omega$ is $q=0$ phonon frequency. 
Since the interlayer stacking arrangement varies spatially, the phonon frequency is influenced by a spatially varying interlayer spring constant~\cite{Wu2014}. 
The weaker van der Waals forces at the AA-stacked region (domain walls), are expected to soften the shear phonon frequency compared to the uniform AB stacked bilayers,  further increasing the effective coupling constant $\lambda$.

\textit{Discussion-} 
Now we return to the possible role of this mechanism in the recently discovered ferroelectric control of superconductivity in bilayer T$_d$-$\text{MoTe}_2$. 
Bulk T$_d$-$\text{MoTe}_2$ is a superconductor but with a very modest   $T_c\approx 120 \text{mK}$. $T_c$ increases as the number of layers is reduced, reaching  $T_c\approx 2 \text{ K}$ in the bilayers and 7 K in the monolayer~\cite{Rhodes2021}. Notably, in the bilayer, $T_c$ is observed to increase sharply at the switching transition of the polarization direction~\cite{Jindal2023}. 
We now discuss the latter property in light of our mechanism. 

When the bilayers are fully polarized in either AB or BA stacking configuration, the shear phonon does not affect polarization to linear order in phonon displacement. (Similarly, the lowest-energy stacking maximizes the interlayer hybridization, eliminating the linear coupling between the shear phonon and hybridization). 
Therefore, it cannot provide significant attractive interaction for superconductivity, leaving only such mechanisms as the standard LO phonons operational. 
Under an applied electric field opposite to the polarization direction, the bilayers start to switch their polarization direction. 
This hysteretic process happens by forming small domains of opposite polarization. 
At the domain walls of these newly formed domains, local pairing becomes enhanced due to the attractive interactions from the domain wall fluctuations. 
However, for weak electric fields, such opposite polarization domains are small and far apart [Fig.~\ref{Fig.:2d_schematic} (a)]. 
Therefore, the ferroelectric phase does not lead to global superconductivity (or enhancement of $T_c$) in transport properties (one could obtain, however, a $local$ enhancement of pairing at the domain walls, possibly detectable with local probes).
As the electric field is increased, the minority domains grow in size and as the system approaches the ferroelectric reversal, the domain walls form a percolating network creating a path for the supercurrent to flow through the system [Fig.~\ref{Fig.:2d_schematic} (b)]. 
This leads to the appearance of superconductivity (or $T_c$ enhancement) in transport properties in this region, as seen in the experiment. 
Finally, as the polarization completely switches to the opposite direction, the superconductivity is again turned off (or $T_c$ is suppressed) with the disappearance of domain walls.

The superconductivity exists in Td-$\text{MoTe}_2$ even away from the ferroelectric switching region. For example, bilayers have $T_c \approx 2.5 K $ at neutrality. 
There need not be a direct relation of this bulk superconductivity to our mechanism. 
However, it also does not stand in contradiction. 
It is possible that another coexisting mechanism is responsible for superconductivity in a single-domain few-layer T$_d$-$\text{MoTe}_2$. In such a case, our mechanism's role is to relatively enhance $T_c$  near the ferroelectric switching.  
A similar assistive role of nematic fluctuations 
in enhancing $T_c$ has been previously studied  in high-temperature superconductors~\cite{Lederer2015,Lederer2017}.

Our mechanism should equally apply to moir\'e bilayers of polar vdW metals. For small twist angles, the lattice relaxation effects lead to large domains of AB and BA regions separated by relatively narrow domain walls, similar to the isolated domain walls that mediate the ferroelectric switching in untwisted bilayers. However, in the moir\'e case, the domain walls automatically form a spanning network, without the need to tune to the switching point. 
The fluctuations of the domain walls would then lead to a spanning superconducting network and global superconductivity.  

When limiting occupancy to the lowest moire subbands, increasing the twist angle may benefit superconductivity, for two reasons. First, that forces the electronic wave-function to enter the domain wall regions (from the AB/BA stacked regions where they are normally peaked) so that they can enjoy stronger electron-phonon interactions.  Also, the larger density of domain walls would lead to a larger average superfluid stiffness.  
We, therefore, anticipate that the peak of $T_c$ in polar moire systems will be reached at the largest twist angles that can still sustain the formation of polar domains.

An observation of superconductivity in twisted bilayers of $\text{MoTe}_2$ can be an indication of the domain wall fluctuation mechanism. 
We are not aware of any such measurement in $\text{MoTe}_2$.
However, the recent observation of superconductivity in twisted bilayers of $\text{WSe}_2$~\cite{Xia2024, Guo2024} provides a possible realization of the domain wall fluctuations mechanism proposed here.

We finally note that since the attractive interactions are enhanced at the domain walls where the polarization vanishes, creating an energetically frustrated {\em spatially uniform} (saddle-point) configuration of $AA$ stacked bilayers of polar vdW metals should provide an ideal setting for such superconductivity. 
In this case, the shear phonons become coupled to the whole system via polar fluctuations (equivalent to the infinite domain wall limit described above). While this would be an unstable configuration in stand-alone bilayers, it is conceivable that such an arrangement can be achieved in artificially stacked multilayers.

\textit{Conclusion-} 
In conclusion, we have shown that domain wall fluctuations in sliding ferroelectric generate local attractive interactions between electrons and can create local Cooper pairs. 
If the domain wall fluctuations originate from phonons, we further express this as an electron-phonon coupling, \textit{albeit} the coupling occurs via a \textit{transverse} piezoelectric mechanism, where induced polarization is proportional to dynamic strain, rather than the much-studied Fr\"oclich coupling between electron charge density and the divergence of lattice polarization (there, the induced polarization is proportional to the LO phonon displacement). 
Further, by considering an interlayer shear phonon as the driver of the polarization fluctuations, we have argued that this coupling can exceed the standard  Fr\"ohlich coupling and is comparable to phonon-interlayer hybridization coupling. 
The proposed scenario for strong electron-phonon coupling is a special feature of the few-layer TMD that are either ordered (or on the verge of ordering) sliding ferroelectrics.
Finally, we have shown that this mechanism may explain the recently discovered polar switching of superconductivity in T$_d$-$\text{MoTe}_2$. 
A full first-principles calculation for electron shear phonon coupling in metallic sliding ferroelectrics will be an interesting direction, which we leave for further studies.

\textit{Acknowledgement-} We thank J. Shi and A. H. MacDonald for useful discussions.
This work was supported by the US Department of Energy, Office of Science, Basic Energy Sciences, Materials Sciences and Engineering Division. 


\bibliography{bibliography}

\begin{thebibliography}{27}%
\makeatletter
\providecommand \@ifxundefined [1]{%
 \@ifx{#1\undefined}
}%
\providecommand \@ifnum [1]{%
 \ifnum #1\expandafter \@firstoftwo
 \else \expandafter \@secondoftwo
 \fi
}%
\providecommand \@ifx [1]{%
 \ifx #1\expandafter \@firstoftwo
 \else \expandafter \@secondoftwo
 \fi
}%
\providecommand \natexlab [1]{#1}%
\providecommand \enquote  [1]{``#1''}%
\providecommand \bibnamefont  [1]{#1}%
\providecommand \bibfnamefont [1]{#1}%
\providecommand \citenamefont [1]{#1}%
\providecommand \href@noop [0]{\@secondoftwo}%
\providecommand \href [0]{\begingroup \@sanitize@url \@href}%
\providecommand \@href[1]{\@@startlink{#1}\@@href}%
\providecommand \@@href[1]{\endgroup#1\@@endlink}%
\providecommand \@sanitize@url [0]{\catcode `\\12\catcode `\$12\catcode `\&12\catcode `\#12\catcode `\^12\catcode `\_12\catcode `\%12\relax}%
\providecommand \@@startlink[1]{}%
\providecommand \@@endlink[0]{}%
\providecommand \url  [0]{\begingroup\@sanitize@url \@url }%
\providecommand \@url [1]{\endgroup\@href {#1}{\urlprefix }}%
\providecommand \urlprefix  [0]{URL }%
\providecommand \Eprint [0]{\href }%
\providecommand \doibase [0]{http://dx.doi.org/}%
\providecommand \selectlanguage [0]{\@gobble}%
\providecommand \bibinfo  [0]{\@secondoftwo}%
\providecommand \bibfield  [0]{\@secondoftwo}%
\providecommand \translation [1]{[#1]}%
\providecommand \BibitemOpen [0]{}%
\providecommand \bibitemStop [0]{}%
\providecommand \bibitemNoStop [0]{.\EOS\space}%
\providecommand \EOS [0]{\spacefactor3000\relax}%
\providecommand \BibitemShut  [1]{\csname bibitem#1\endcsname}%
\let\auto@bib@innerbib\@empty
\bibitem [{\citenamefont {Fei}\ \emph {et~al.}(2018)\citenamefont {Fei}, \citenamefont {Zhao}, \citenamefont {Palomaki}, \citenamefont {Sun}, \citenamefont {Miller}, \citenamefont {Zhao}, \citenamefont {Yan}, \citenamefont {Xu},\ and\ \citenamefont {Cobden}}]{Fei2018}%
  \BibitemOpen
  \bibfield  {author} {\bibinfo {author} {\bibfnamefont {Z.}~\bibnamefont {Fei}}, \bibinfo {author} {\bibfnamefont {W.}~\bibnamefont {Zhao}}, \bibinfo {author} {\bibfnamefont {T.~A.}\ \bibnamefont {Palomaki}}, \bibinfo {author} {\bibfnamefont {B.}~\bibnamefont {Sun}}, \bibinfo {author} {\bibfnamefont {M.~K.}\ \bibnamefont {Miller}}, \bibinfo {author} {\bibfnamefont {Z.}~\bibnamefont {Zhao}}, \bibinfo {author} {\bibfnamefont {J.}~\bibnamefont {Yan}}, \bibinfo {author} {\bibfnamefont {X.}~\bibnamefont {Xu}}, \ and\ \bibinfo {author} {\bibfnamefont {D.~H.}\ \bibnamefont {Cobden}},\ }\bibfield  {title} {\enquote {\bibinfo {title} {Ferroelectric switching of a two-dimensional metal},}\ }\href {\doibase 10.1038/s41586-018-0336-3} {\bibfield  {journal} {\bibinfo  {journal} {Nature}\ }\textbf {\bibinfo {volume} {560}},\ \bibinfo {pages} {336} (\bibinfo {year} {2018})}\BibitemShut {NoStop}%
\bibitem [{\citenamefont {Yasuda}\ \emph {et~al.}(2021)\citenamefont {Yasuda}, \citenamefont {Wang}, \citenamefont {Watanabe}, \citenamefont {Taniguchi},\ and\ \citenamefont {Jarillo-Herrero}}]{Yasuda2021}%
  \BibitemOpen
  \bibfield  {author} {\bibinfo {author} {\bibfnamefont {K.}~\bibnamefont {Yasuda}}, \bibinfo {author} {\bibfnamefont {X.}~\bibnamefont {Wang}}, \bibinfo {author} {\bibfnamefont {K.}~\bibnamefont {Watanabe}}, \bibinfo {author} {\bibfnamefont {T.}~\bibnamefont {Taniguchi}}, \ and\ \bibinfo {author} {\bibfnamefont {P.}~\bibnamefont {Jarillo-Herrero}},\ }\bibfield  {title} {\enquote {\bibinfo {title} {Stacking-engineered ferroelectricity in bilayer boron nitride},}\ }\href {\doibase 10.1126/science.abd3230} {\bibfield  {journal} {\bibinfo  {journal} {Science}\ }\textbf {\bibinfo {volume} {372}},\ \bibinfo {pages} {1458} (\bibinfo {year} {2021})}\BibitemShut {NoStop}%
\bibitem [{\citenamefont {Stern}\ \emph {et~al.}(2021)\citenamefont {Stern}, \citenamefont {Waschitz}, \citenamefont {Cao}, \citenamefont {Nevo}, \citenamefont {Watanabe}, \citenamefont {Taniguchi}, \citenamefont {Sela}, \citenamefont {Urbakh}, \citenamefont {Hod},\ and\ \citenamefont {Shalom}}]{Stern2021}%
  \BibitemOpen
  \bibfield  {author} {\bibinfo {author} {\bibfnamefont {M.~V.}\ \bibnamefont {Stern}}, \bibinfo {author} {\bibfnamefont {Y.}~\bibnamefont {Waschitz}}, \bibinfo {author} {\bibfnamefont {W.}~\bibnamefont {Cao}}, \bibinfo {author} {\bibfnamefont {I.}~\bibnamefont {Nevo}}, \bibinfo {author} {\bibfnamefont {K.}~\bibnamefont {Watanabe}}, \bibinfo {author} {\bibfnamefont {T.}~\bibnamefont {Taniguchi}}, \bibinfo {author} {\bibfnamefont {E.}~\bibnamefont {Sela}}, \bibinfo {author} {\bibfnamefont {M.}~\bibnamefont {Urbakh}}, \bibinfo {author} {\bibfnamefont {O.}~\bibnamefont {Hod}}, \ and\ \bibinfo {author} {\bibfnamefont {M.~B.}\ \bibnamefont {Shalom}},\ }\bibfield  {title} {\enquote {\bibinfo {title} {Interfacial ferroelectricity by van der waals sliding},}\ }\href {\doibase 10.1126/science.abe8177} {\bibfield  {journal} {\bibinfo  {journal} {Science}\ }\textbf {\bibinfo {volume} {372}},\ \bibinfo {pages} {1462} (\bibinfo {year} {2021})}\BibitemShut {NoStop}%
\bibitem [{\citenamefont {de~la Barrera}\ \emph {et~al.}(2021)\citenamefont {de~la Barrera}, \citenamefont {Cao}, \citenamefont {Gao}, \citenamefont {Gao}, \citenamefont {Bheemarasetty}, \citenamefont {Yan}, \citenamefont {Mandrus}, \citenamefont {Zhu}, \citenamefont {Xiao},\ and\ \citenamefont {Hunt}}]{Barrera2021}%
  \BibitemOpen
  \bibfield  {author} {\bibinfo {author} {\bibfnamefont {S.~C.}\ \bibnamefont {de~la Barrera}}, \bibinfo {author} {\bibfnamefont {Q.}~\bibnamefont {Cao}}, \bibinfo {author} {\bibfnamefont {Y.}~\bibnamefont {Gao}}, \bibinfo {author} {\bibfnamefont {Y.}~\bibnamefont {Gao}}, \bibinfo {author} {\bibfnamefont {V.~S.}\ \bibnamefont {Bheemarasetty}}, \bibinfo {author} {\bibfnamefont {J.}~\bibnamefont {Yan}}, \bibinfo {author} {\bibfnamefont {D.~G.}\ \bibnamefont {Mandrus}}, \bibinfo {author} {\bibfnamefont {W.}~\bibnamefont {Zhu}}, \bibinfo {author} {\bibfnamefont {D.}~\bibnamefont {Xiao}}, \ and\ \bibinfo {author} {\bibfnamefont {B.~M.}\ \bibnamefont {Hunt}},\ }\bibfield  {title} {\enquote {\bibinfo {title} {Direct measurement of ferroelectric polarization in a tunable semimetal},}\ }\href {\doibase 10.1038/s41467-021-25587-3} {\bibfield  {journal} {\bibinfo  {journal} {Nat. Commun.}\ }\textbf {\bibinfo {volume} {12}},\ \bibinfo {pages} {5298} (\bibinfo {year} {2021})}\BibitemShut {NoStop}%
\bibitem [{\citenamefont {Wang}\ \emph {et~al.}(2022)\citenamefont {Wang}, \citenamefont {Yasuda}, \citenamefont {Zhang}, \citenamefont {Liu}, \citenamefont {Watanabe}, \citenamefont {Taniguchi}, \citenamefont {Hone}, \citenamefont {Fu},\ and\ \citenamefont {Jarillo-Herrero}}]{Wang2022}%
  \BibitemOpen
  \bibfield  {author} {\bibinfo {author} {\bibfnamefont {X.}~\bibnamefont {Wang}}, \bibinfo {author} {\bibfnamefont {K.}~\bibnamefont {Yasuda}}, \bibinfo {author} {\bibfnamefont {Y.}~\bibnamefont {Zhang}}, \bibinfo {author} {\bibfnamefont {S.}~\bibnamefont {Liu}}, \bibinfo {author} {\bibfnamefont {K.}~\bibnamefont {Watanabe}}, \bibinfo {author} {\bibfnamefont {T.}~\bibnamefont {Taniguchi}}, \bibinfo {author} {\bibfnamefont {J.}~\bibnamefont {Hone}}, \bibinfo {author} {\bibfnamefont {L.}~\bibnamefont {Fu}}, \ and\ \bibinfo {author} {\bibfnamefont {P.}~\bibnamefont {Jarillo-Herrero}},\ }\bibfield  {title} {\enquote {\bibinfo {title} {Interfacial ferroelectricity in rhombohedral-stacked bilayer transition metal dichalcogenides},}\ }\href {\doibase 10.1038/s41565-021-01059-z} {\bibfield  {journal} {\bibinfo  {journal} {Nat. Nanotech.}\ }\textbf {\bibinfo {volume} {17}},\ \bibinfo {pages} {367} (\bibinfo {year} {2022})}\BibitemShut {NoStop}%
\bibitem [{\citenamefont {Enaldiev}\ \emph {et~al.}(2021)\citenamefont {Enaldiev}, \citenamefont {Ferreira}, \citenamefont {Magorrian},\ and\ \citenamefont {Fal’ko}}]{Enaldiev2021}%
  \BibitemOpen
  \bibfield  {author} {\bibinfo {author} {\bibfnamefont {V.~V.}\ \bibnamefont {Enaldiev}}, \bibinfo {author} {\bibfnamefont {F.}~\bibnamefont {Ferreira}}, \bibinfo {author} {\bibfnamefont {S.~J.}\ \bibnamefont {Magorrian}}, \ and\ \bibinfo {author} {\bibfnamefont {V.~I.}\ \bibnamefont {Fal’ko}},\ }\bibfield  {title} {\enquote {\bibinfo {title} {Piezoelectric networks and ferroelectric domains in twistronic superlattices in ws2/mos2 and wse2/mose2 bilayers},}\ }\href {\doibase 10.1088/2053-1583/abdd92} {\bibfield  {journal} {\bibinfo  {journal} {2D Mater.}\ }\textbf {\bibinfo {volume} {8}},\ \bibinfo {pages} {025030} (\bibinfo {year} {2021})}\BibitemShut {NoStop}%
\bibitem [{\citenamefont {Enaldiev}\ \emph {et~al.}(2022)\citenamefont {Enaldiev}, \citenamefont {Ferreira},\ and\ \citenamefont {Fal’ko}}]{Enaldiev2022}%
  \BibitemOpen
  \bibfield  {author} {\bibinfo {author} {\bibfnamefont {V.~V.}\ \bibnamefont {Enaldiev}}, \bibinfo {author} {\bibfnamefont {F.}~\bibnamefont {Ferreira}}, \ and\ \bibinfo {author} {\bibfnamefont {V.~I.}\ \bibnamefont {Fal’ko}},\ }\bibfield  {title} {\enquote {\bibinfo {title} {A scalable network model for electrically tunable ferroelectric domain structure in twistronic bilayers of two-dimensional semiconductors},}\ }\href {\doibase 10.1021/acs.nanolett.1c04210} {\bibfield  {journal} {\bibinfo  {journal} {Nano Letters}\ }\textbf {\bibinfo {volume} {22}},\ \bibinfo {pages} {1534} (\bibinfo {year} {2022})}\BibitemShut {NoStop}%
\bibitem [{\citenamefont {Bennett}\ and\ \citenamefont {Remez}(2022)}]{Bennett2022}%
  \BibitemOpen
  \bibfield  {author} {\bibinfo {author} {\bibfnamefont {D.}~\bibnamefont {Bennett}}\ and\ \bibinfo {author} {\bibfnamefont {B.}~\bibnamefont {Remez}},\ }\bibfield  {title} {\enquote {\bibinfo {title} {On electrically tunable stacking domains and ferroelectricity in moiré superlattices},}\ }\href {\doibase 10.1038/s41699-021-00281-6} {\bibfield  {journal} {\bibinfo  {journal} {npj 2D Mater. Appl.}\ }\textbf {\bibinfo {volume} {6}},\ \bibinfo {pages} {1} (\bibinfo {year} {2022})}\BibitemShut {NoStop}%
\bibitem [{\citenamefont {Bennett}\ \emph {et~al.}(2023)\citenamefont {Bennett}, \citenamefont {Chaudhary}, \citenamefont {Slager}, \citenamefont {Bousquet},\ and\ \citenamefont {Ghosez}}]{Bennett2023}%
  \BibitemOpen
  \bibfield  {author} {\bibinfo {author} {\bibfnamefont {D.}~\bibnamefont {Bennett}}, \bibinfo {author} {\bibfnamefont {G.}~\bibnamefont {Chaudhary}}, \bibinfo {author} {\bibfnamefont {R.-J.}\ \bibnamefont {Slager}}, \bibinfo {author} {\bibfnamefont {E.}~\bibnamefont {Bousquet}}, \ and\ \bibinfo {author} {\bibfnamefont {P.}~\bibnamefont {Ghosez}},\ }\bibfield  {title} {\enquote {\bibinfo {title} {Polar meron-antimeron networks in strained and twisted bilayers},}\ }\href {\doibase 10.1038/s41467-023-37337-8} {\bibfield  {journal} {\bibinfo  {journal} {Nat. Commun.}\ }\textbf {\bibinfo {volume} {14}},\ \bibinfo {pages} {1629} (\bibinfo {year} {2023})}\BibitemShut {NoStop}%
\bibitem [{\citenamefont {Zheng}\ \emph {et~al.}(2020)\citenamefont {Zheng}, \citenamefont {Ma}, \citenamefont {Bi}, \citenamefont {de~la Barrera}, \citenamefont {Liu}, \citenamefont {Mao}, \citenamefont {Zhang}, \citenamefont {Kiper}, \citenamefont {Watanabe}, \citenamefont {Taniguchi}, \citenamefont {Kong}, \citenamefont {Tisdale}, \citenamefont {Ashoori}, \citenamefont {Gedik}, \citenamefont {Fu}, \citenamefont {Xu},\ and\ \citenamefont {Jarillo.-Herrero}}]{Zheng2020}%
  \BibitemOpen
  \bibfield  {author} {\bibinfo {author} {\bibfnamefont {Z.}~\bibnamefont {Zheng}}, \bibinfo {author} {\bibfnamefont {Q.}~\bibnamefont {Ma}}, \bibinfo {author} {\bibfnamefont {Z.}~\bibnamefont {Bi}}, \bibinfo {author} {\bibfnamefont {S.}~\bibnamefont {de~la Barrera}}, \bibinfo {author} {\bibfnamefont {M.-H.}\ \bibnamefont {Liu}}, \bibinfo {author} {\bibfnamefont {N.}~\bibnamefont {Mao}}, \bibinfo {author} {\bibfnamefont {Y.}~\bibnamefont {Zhang}}, \bibinfo {author} {\bibfnamefont {N.}~\bibnamefont {Kiper}}, \bibinfo {author} {\bibfnamefont {K.}~\bibnamefont {Watanabe}}, \bibinfo {author} {\bibfnamefont {T.}~\bibnamefont {Taniguchi}}, \bibinfo {author} {\bibfnamefont {J.}~\bibnamefont {Kong}}, \bibinfo {author} {\bibfnamefont {W.~A.}\ \bibnamefont {Tisdale}}, \bibinfo {author} {\bibfnamefont {R.}~\bibnamefont {Ashoori}}, \bibinfo {author} {\bibfnamefont {N.}~\bibnamefont {Gedik}}, \bibinfo {author} {\bibfnamefont {L.}~\bibnamefont {Fu}}, \bibinfo {author} {\bibfnamefont {S.-Y.}\ \bibnamefont {Xu}}, \ and\
  \bibinfo {author} {\bibfnamefont {P.}~\bibnamefont {Jarillo.-Herrero}},\ }\bibfield  {title} {\enquote {\bibinfo {title} {Unconventional ferroelectricity in moiré heterostructures},}\ }\href {\doibase 10.1038/s41586-020-2970-9} {\bibfield  {journal} {\bibinfo  {journal} {Nature}\ }\textbf {\bibinfo {volume} {588}},\ \bibinfo {pages} {71} (\bibinfo {year} {2020})}\BibitemShut {NoStop}%
\bibitem [{\citenamefont {Niu}\ \emph {et~al.}(2022)\citenamefont {Niu}, \citenamefont {Li}, \citenamefont {Han}, \citenamefont {Qu}, \citenamefont {Ding}, \citenamefont {Wang}, \citenamefont {Liu}, \citenamefont {Liu}, \citenamefont {Han}, \citenamefont {Watanabe}, \citenamefont {Taniguchi}, \citenamefont {Wu}, \citenamefont {Ren}, \citenamefont {Wang}, \citenamefont {Hong}, \citenamefont {Mao}, \citenamefont {Han}, \citenamefont {Liu}, \citenamefont {Gan},\ and\ \citenamefont {Lu}}]{Niu2022}%
  \BibitemOpen
  \bibfield  {author} {\bibinfo {author} {\bibfnamefont {R.}~\bibnamefont {Niu}}, \bibinfo {author} {\bibfnamefont {Z.}~\bibnamefont {Li}}, \bibinfo {author} {\bibfnamefont {X.}~\bibnamefont {Han}}, \bibinfo {author} {\bibfnamefont {Z.}~\bibnamefont {Qu}}, \bibinfo {author} {\bibfnamefont {D.}~\bibnamefont {Ding}}, \bibinfo {author} {\bibfnamefont {Z.}~\bibnamefont {Wang}}, \bibinfo {author} {\bibfnamefont {Q.}~\bibnamefont {Liu}}, \bibinfo {author} {\bibfnamefont {T.}~\bibnamefont {Liu}}, \bibinfo {author} {\bibfnamefont {C.}~\bibnamefont {Han}}, \bibinfo {author} {\bibfnamefont {K.}~\bibnamefont {Watanabe}}, \bibinfo {author} {\bibfnamefont {T.}~\bibnamefont {Taniguchi}}, \bibinfo {author} {\bibfnamefont {M.}~\bibnamefont {Wu}}, \bibinfo {author} {\bibfnamefont {Q.}~\bibnamefont {Ren}}, \bibinfo {author} {\bibfnamefont {X.}~\bibnamefont {Wang}}, \bibinfo {author} {\bibfnamefont {J.}~\bibnamefont {Hong}}, \bibinfo {author} {\bibfnamefont {J.}~\bibnamefont {Mao}}, \bibinfo {author} {\bibfnamefont
  {Z.}~\bibnamefont {Han}}, \bibinfo {author} {\bibfnamefont {K.}~\bibnamefont {Liu}}, \bibinfo {author} {\bibfnamefont {Z.}~\bibnamefont {Gan}}, \ and\ \bibinfo {author} {\bibfnamefont {J.}~\bibnamefont {Lu}},\ }\bibfield  {title} {\enquote {\bibinfo {title} {Giant ferroelectric polarization in a bilayer graphene heterostructure},}\ }\href {\doibase 10.1038/s41467-022-34104-z} {\bibfield  {journal} {\bibinfo  {journal} {Nat. Commun.}\ }\textbf {\bibinfo {volume} {13}},\ \bibinfo {pages} {6241} (\bibinfo {year} {2022})}\BibitemShut {NoStop}%
\bibitem [{\citenamefont {Zheng}\ \emph {et~al.}()\citenamefont {Zheng}, \citenamefont {Wang}, \citenamefont {Zhu}, \citenamefont {Carr}, \citenamefont {Devakul}, \citenamefont {de~la Barrera}, \citenamefont {Paul}, \citenamefont {Huang}, \citenamefont {Gao}, \citenamefont {Zhang}, \citenamefont {Bérubé}, \citenamefont {Evancho}, \citenamefont {Watanabe}, \citenamefont {Taniguchi}, \citenamefont {Fu}, \citenamefont {Wang}, \citenamefont {Xu}, \citenamefont {Kaxiras}, \citenamefont {Jarillo-Herrero},\ and\ \citenamefont {Ma}}]{Zheng2023}%
  \BibitemOpen
  \bibfield  {author} {\bibinfo {author} {\bibfnamefont {Z}~\bibnamefont {Zheng}}, \bibinfo {author} {\bibfnamefont {X.}~\bibnamefont {Wang}}, \bibinfo {author} {\bibfnamefont {Z.}~\bibnamefont {Zhu}}, \bibinfo {author} {\bibfnamefont {S.}~\bibnamefont {Carr}}, \bibinfo {author} {\bibfnamefont {T.}~\bibnamefont {Devakul}}, \bibinfo {author} {\bibfnamefont {S.}~\bibnamefont {de~la Barrera}}, \bibinfo {author} {\bibfnamefont {N.}~\bibnamefont {Paul}}, \bibinfo {author} {\bibfnamefont {Z.}~\bibnamefont {Huang}}, \bibinfo {author} {\bibfnamefont {A.}~\bibnamefont {Gao}}, \bibinfo {author} {\bibfnamefont {Y.}~\bibnamefont {Zhang}}, \bibinfo {author} {\bibfnamefont {D.}~\bibnamefont {Bérubé}}, \bibinfo {author} {\bibfnamefont {K.~N.}\ \bibnamefont {Evancho}}, \bibinfo {author} {\bibfnamefont {K.}~\bibnamefont {Watanabe}}, \bibinfo {author} {\bibfnamefont {T.}~\bibnamefont {Taniguchi}}, \bibinfo {author} {\bibfnamefont {L.}~\bibnamefont {Fu}}, \bibinfo {author} {\bibfnamefont {Y.}~\bibnamefont {Wang}}, \bibinfo
  {author} {\bibfnamefont {S.-Y.}\ \bibnamefont {Xu}}, \bibinfo {author} {\bibfnamefont {E.}~\bibnamefont {Kaxiras}}, \bibinfo {author} {\bibfnamefont {P.}~\bibnamefont {Jarillo-Herrero}}, \ and\ \bibinfo {author} {\bibfnamefont {Q.}~\bibnamefont {Ma}},\ }\bibfield  {title} {\enquote {\bibinfo {title} {Electronic ratchet effect in a moiré system: signatures of excitonic ferroelectricity},}\ }\href@noop {} {\ }\Eprint {http://arxiv.org/abs/arXiv:2306.03922} {arXiv:2306.03922} \BibitemShut {NoStop}%
\bibitem [{\citenamefont {Jindal}\ \emph {et~al.}(2023)\citenamefont {Jindal}, \citenamefont {Saha}, \citenamefont {Li}, \citenamefont {Taniguchi}, \citenamefont {Watanabe}, \citenamefont {Hone}, \citenamefont {Birol}, \citenamefont {Fernandes}, \citenamefont {Dean}, \citenamefont {Pasupathy},\ and\ \citenamefont {Rhodes}}]{Jindal2023}%
  \BibitemOpen
  \bibfield  {author} {\bibinfo {author} {\bibfnamefont {A.}~\bibnamefont {Jindal}}, \bibinfo {author} {\bibfnamefont {A.}~\bibnamefont {Saha}}, \bibinfo {author} {\bibfnamefont {Z.}~\bibnamefont {Li}}, \bibinfo {author} {\bibfnamefont {T.}~\bibnamefont {Taniguchi}}, \bibinfo {author} {\bibfnamefont {K.}~\bibnamefont {Watanabe}}, \bibinfo {author} {\bibfnamefont {J.~C.}\ \bibnamefont {Hone}}, \bibinfo {author} {\bibfnamefont {T.}~\bibnamefont {Birol}}, \bibinfo {author} {\bibfnamefont {R.~M.}\ \bibnamefont {Fernandes}}, \bibinfo {author} {\bibfnamefont {C.~R.}\ \bibnamefont {Dean}}, \bibinfo {author} {\bibfnamefont {A.~N.}\ \bibnamefont {Pasupathy}}, \ and\ \bibinfo {author} {\bibfnamefont {D.~A.}\ \bibnamefont {Rhodes}},\ }\bibfield  {title} {\enquote {\bibinfo {title} {Coupled ferroelectricity and superconductivity in bilayer td-mote2},}\ }\href {\doibase 10.1038/s41586-022-05521-3} {\bibfield  {journal} {\bibinfo  {journal} {Nature}\ }\textbf {\bibinfo {volume} {613}},\ \bibinfo {pages} {48} (\bibinfo
  {year} {2023})}\BibitemShut {NoStop}%
\bibitem [{\citenamefont {Bardeen}\ \emph {et~al.}(1957)\citenamefont {Bardeen}, \citenamefont {Cooper},\ and\ \citenamefont {Schrieffer}}]{Bardeen1957}%
  \BibitemOpen
  \bibfield  {author} {\bibinfo {author} {\bibfnamefont {J.}~\bibnamefont {Bardeen}}, \bibinfo {author} {\bibfnamefont {L.~N.}\ \bibnamefont {Cooper}}, \ and\ \bibinfo {author} {\bibfnamefont {J.~R.}\ \bibnamefont {Schrieffer}},\ }\bibfield  {title} {\enquote {\bibinfo {title} {Theory of superconductivity},}\ }\href {\doibase 10.1103/PhysRev.108.1175} {\bibfield  {journal} {\bibinfo  {journal} {Phys. Rev.}\ }\textbf {\bibinfo {volume} {108}},\ \bibinfo {pages} {1175} (\bibinfo {year} {1957})}\BibitemShut {NoStop}%
\bibitem [{\citenamefont {Moriya}\ \emph {et~al.}(1990)\citenamefont {Moriya}, \citenamefont {Takahashi},\ and\ \citenamefont {Ueda}}]{Moriya1990}%
  \BibitemOpen
  \bibfield  {author} {\bibinfo {author} {\bibfnamefont {T.}~\bibnamefont {Moriya}}, \bibinfo {author} {\bibfnamefont {Y.}~\bibnamefont {Takahashi}}, \ and\ \bibinfo {author} {\bibfnamefont {K.}~\bibnamefont {Ueda}},\ }\bibfield  {title} {\enquote {\bibinfo {title} {Antiferromagnetic spin fluctuations and superconductivity in two-dimensional metals -a possible model for high tc oxides},}\ }\href {\doibase 10.1143/JPSJ.59.2905} {\bibfield  {journal} {\bibinfo  {journal} {JPSJ}\ }\textbf {\bibinfo {volume} {59}},\ \bibinfo {pages} {2905} (\bibinfo {year} {1990})}\BibitemShut {NoStop}%
\bibitem [{\citenamefont {Mahan}(2000)}]{mahan2000many}%
  \BibitemOpen
  \bibfield  {author} {\bibinfo {author} {\bibfnamefont {GD}~\bibnamefont {Mahan}},\ }\href@noop {} {\enquote {\bibinfo {title} {Many-body physics},}\ } (\bibinfo {year} {2000})\BibitemShut {NoStop}%
\bibitem [{\citenamefont {Martin}\ \emph {et~al.}(2005)\citenamefont {Martin}, \citenamefont {Podolsky},\ and\ \citenamefont {Kivelson}}]{PhysRevB.72.060502}%
  \BibitemOpen
  \bibfield  {author} {\bibinfo {author} {\bibfnamefont {Ivar}\ \bibnamefont {Martin}}, \bibinfo {author} {\bibfnamefont {Daniel}\ \bibnamefont {Podolsky}}, \ and\ \bibinfo {author} {\bibfnamefont {Steven~A.}\ \bibnamefont {Kivelson}},\ }\bibfield  {title} {\enquote {\bibinfo {title} {Enhancement of superconductivity by local inhomogeneities},}\ }\href {\doibase 10.1103/PhysRevB.72.060502} {\bibfield  {journal} {\bibinfo  {journal} {Phys. Rev. B}\ }\textbf {\bibinfo {volume} {72}},\ \bibinfo {pages} {060502} (\bibinfo {year} {2005})}\BibitemShut {NoStop}%
\bibitem [{\citenamefont {Fr{\"o}hlich}(1954)}]{frohlich1954electrons}%
  \BibitemOpen
  \bibfield  {author} {\bibinfo {author} {\bibfnamefont {Herbert}\ \bibnamefont {Fr{\"o}hlich}},\ }\bibfield  {title} {\enquote {\bibinfo {title} {Electrons in lattice fields},}\ }\href@noop {} {\bibfield  {journal} {\bibinfo  {journal} {Advances in Physics}\ }\textbf {\bibinfo {volume} {3}},\ \bibinfo {pages} {325--361} (\bibinfo {year} {1954})}\BibitemShut {NoStop}%
\bibitem [{\citenamefont {Wu}\ and\ \citenamefont {Ju}(2021)}]{Wu2021}%
  \BibitemOpen
  \bibfield  {author} {\bibinfo {author} {\bibfnamefont {M.}~\bibnamefont {Wu}}\ and\ \bibinfo {author} {\bibfnamefont {L.}~\bibnamefont {Ju}},\ }\bibfield  {title} {\enquote {\bibinfo {title} {Sliding ferroelectricity in 2d van der waals materials: Related physics and future opportunities},}\ }\href {\doibase 10.1073/pnas.2115703118} {\bibfield  {journal} {\bibinfo  {journal} {PNAS}\ }\textbf {\bibinfo {volume} {118}},\ \bibinfo {pages} {e2115703118} (\bibinfo {year} {2021})}\BibitemShut {NoStop}%
\bibitem [{\citenamefont {McMillan}(1968)}]{McMillan1968}%
  \BibitemOpen
  \bibfield  {author} {\bibinfo {author} {\bibfnamefont {W.~L.}\ \bibnamefont {McMillan}},\ }\bibfield  {title} {\enquote {\bibinfo {title} {Transition temperature of strong-coupled superconductors},}\ }\href {\doibase 10.1103/PhysRev.167.331} {\bibfield  {journal} {\bibinfo  {journal} {Phys. Rev.}\ }\textbf {\bibinfo {volume} {167}},\ \bibinfo {pages} {331} (\bibinfo {year} {1968})}\BibitemShut {NoStop}%
\bibitem [{Note1()}]{Note1}%
  \BibitemOpen
  \bibinfo {note} {The $q=0$ frequency of the shear phonon is determined by the weak interlayer spring constant. However, its dispersion is predominantly governed by the stiff intralayer spring constant. Therefore, these phonons are highly dispersive and neglecting their dispersion is not a very good approximation.}\BibitemShut {Stop}%
\bibitem [{\citenamefont {Wu}\ \emph {et~al.}(2014)\citenamefont {Wu}, \citenamefont {Zhang}, \citenamefont {Ijäs}, \citenamefont {Han}, \citenamefont {Qiao}, \citenamefont {Li}, \citenamefont {Jiang}, \citenamefont {Ferrari},\ and\ \citenamefont {Tan}}]{Wu2014}%
  \BibitemOpen
  \bibfield  {author} {\bibinfo {author} {\bibfnamefont {J.-B.}\ \bibnamefont {Wu}}, \bibinfo {author} {\bibfnamefont {X.}~\bibnamefont {Zhang}}, \bibinfo {author} {\bibfnamefont {M.}~\bibnamefont {Ijäs}}, \bibinfo {author} {\bibfnamefont {W.-P.}\ \bibnamefont {Han}}, \bibinfo {author} {\bibfnamefont {X.-F.}\ \bibnamefont {Qiao}}, \bibinfo {author} {\bibfnamefont {X.-L.}\ \bibnamefont {Li}}, \bibinfo {author} {\bibfnamefont {D.-S.}\ \bibnamefont {Jiang}}, \bibinfo {author} {\bibfnamefont {A.~C.}\ \bibnamefont {Ferrari}}, \ and\ \bibinfo {author} {\bibfnamefont {P.-H.}\ \bibnamefont {Tan}},\ }\bibfield  {title} {\enquote {\bibinfo {title} {Resonant raman spectroscopy of twisted multilayer graphene},}\ }\href {\doibase 10.1038/ncomms6309} {\bibfield  {journal} {\bibinfo  {journal} {Nat. Commun.}\ }\textbf {\bibinfo {volume} {5}},\ \bibinfo {pages} {5309} (\bibinfo {year} {2014})}\BibitemShut {NoStop}%
\bibitem [{\citenamefont {Rhodes}\ \emph {et~al.}(2021)\citenamefont {Rhodes}, \citenamefont {Jindal}, \citenamefont {Yuan}, \citenamefont {Jung}, \citenamefont {Antony}, \citenamefont {Wang}, \citenamefont {Kim}, \citenamefont {Chiu}, \citenamefont {Taniguchi}, \citenamefont {Watanabe}, \citenamefont {Barmak}, \citenamefont {Balicas}, \citenamefont {Dean}, \citenamefont {Qian}, \citenamefont {Fu}, \citenamefont {Pasupathy},\ and\ \citenamefont {Hone}}]{Rhodes2021}%
  \BibitemOpen
  \bibfield  {author} {\bibinfo {author} {\bibfnamefont {D.~A.}\ \bibnamefont {Rhodes}}, \bibinfo {author} {\bibfnamefont {A.}~\bibnamefont {Jindal}}, \bibinfo {author} {\bibfnamefont {N.~F.~Q.}\ \bibnamefont {Yuan}}, \bibinfo {author} {\bibfnamefont {Y.}~\bibnamefont {Jung}}, \bibinfo {author} {\bibfnamefont {A.}~\bibnamefont {Antony}}, \bibinfo {author} {\bibfnamefont {H.}~\bibnamefont {Wang}}, \bibinfo {author} {\bibfnamefont {B.}~\bibnamefont {Kim}}, \bibinfo {author} {\bibfnamefont {Y.~C.}\ \bibnamefont {Chiu}}, \bibinfo {author} {\bibfnamefont {T.}~\bibnamefont {Taniguchi}}, \bibinfo {author} {\bibfnamefont {K.}~\bibnamefont {Watanabe}}, \bibinfo {author} {\bibfnamefont {K.}~\bibnamefont {Barmak}}, \bibinfo {author} {\bibfnamefont {L.}~\bibnamefont {Balicas}}, \bibinfo {author} {\bibfnamefont {C.~R.}\ \bibnamefont {Dean}}, \bibinfo {author} {\bibfnamefont {X.}~\bibnamefont {Qian}}, \bibinfo {author} {\bibfnamefont {L.}~\bibnamefont {Fu}}, \bibinfo {author} {\bibfnamefont {A.~N.}\ \bibnamefont
  {Pasupathy}}, \ and\ \bibinfo {author} {\bibfnamefont {J.}~\bibnamefont {Hone}},\ }\bibfield  {title} {\enquote {\bibinfo {title} {Superconductivity in monolayer td-mote2},}\ }\href {\doibase 10.1021/acs.nanolett.0c04935} {\bibfield  {journal} {\bibinfo  {journal} {Nano Lett.}\ ,\ \bibinfo {pages} {2505}} (\bibinfo {year} {2021})}\BibitemShut {NoStop}%
\bibitem [{\citenamefont {Lederer}\ \emph {et~al.}(2015)\citenamefont {Lederer}, \citenamefont {Schattner}, \citenamefont {Berg},\ and\ \citenamefont {Kivelson}}]{Lederer2015}%
  \BibitemOpen
  \bibfield  {author} {\bibinfo {author} {\bibfnamefont {S.}~\bibnamefont {Lederer}}, \bibinfo {author} {\bibfnamefont {Y.}~\bibnamefont {Schattner}}, \bibinfo {author} {\bibfnamefont {E.}~\bibnamefont {Berg}}, \ and\ \bibinfo {author} {\bibfnamefont {S.~A.}\ \bibnamefont {Kivelson}},\ }\bibfield  {title} {\enquote {\bibinfo {title} {Enhancement of superconductivity near a nematic quantum critical point},}\ }\href {\doibase 10.1103/PhysRevLett.114.097001} {\bibfield  {journal} {\bibinfo  {journal} {Phys. Rev. Lett.}\ }\textbf {\bibinfo {volume} {114}},\ \bibinfo {pages} {097001} (\bibinfo {year} {2015})}\BibitemShut {NoStop}%
\bibitem [{\citenamefont {Lederer}\ \emph {et~al.}(2017)\citenamefont {Lederer}, \citenamefont {Schattner}, \citenamefont {Berg},\ and\ \citenamefont {Kivelson}}]{Lederer2017}%
  \BibitemOpen
  \bibfield  {author} {\bibinfo {author} {\bibfnamefont {S.}~\bibnamefont {Lederer}}, \bibinfo {author} {\bibfnamefont {Y.}~\bibnamefont {Schattner}}, \bibinfo {author} {\bibfnamefont {E.}~\bibnamefont {Berg}}, \ and\ \bibinfo {author} {\bibfnamefont {S.~A.}\ \bibnamefont {Kivelson}},\ }\bibfield  {title} {\enquote {\bibinfo {title} {Superconductivity and non-fermi liquid behavior near a nematic quantum critical point},}\ }\href {\doibase 10.1073/pnas.1620651114} {\bibfield  {journal} {\bibinfo  {journal} {PNAS}\ }\textbf {\bibinfo {volume} {114}},\ \bibinfo {pages} {4905} (\bibinfo {year} {2017})}\BibitemShut {NoStop}%
\bibitem [{\citenamefont {Xia}\ \emph {et~al.}()\citenamefont {Xia}, \citenamefont {Han}, \citenamefont {Watanabe}, \citenamefont {Taniguchi}, \citenamefont {Shan},\ and\ \citenamefont {Mak}}]{Xia2024}%
  \BibitemOpen
  \bibfield  {author} {\bibinfo {author} {\bibfnamefont {Y.}~\bibnamefont {Xia}}, \bibinfo {author} {\bibfnamefont {Z.}~\bibnamefont {Han}}, \bibinfo {author} {\bibfnamefont {K.}~\bibnamefont {Watanabe}}, \bibinfo {author} {\bibfnamefont {T.}~\bibnamefont {Taniguchi}}, \bibinfo {author} {\bibfnamefont {J.}~\bibnamefont {Shan}}, \ and\ \bibinfo {author} {\bibfnamefont {K.~F.}\ \bibnamefont {Mak}},\ }\bibfield  {title} {\enquote {\bibinfo {title} {Unconventional superconductivity in twisted bilayer wse2},}\ }\href@noop {} {\ }\Eprint {http://arxiv.org/abs/arXiv:2405.14784} {arXiv:2405.14784} \BibitemShut {NoStop}%
\bibitem [{\citenamefont {Guo}\ \emph {et~al.}()\citenamefont {Guo}, \citenamefont {Pack}, \citenamefont {Swann}, \citenamefont {Holtzman}, \citenamefont {Cothrine}, \citenamefont {Watanabe}, \citenamefont {Taniguchi}, \citenamefont {Mandrus}, \citenamefont {Barmak}, \citenamefont {Hone}, \citenamefont {Millis}, \citenamefont {Pasupathy},\ and\ \citenamefont {Dean}}]{Guo2024}%
  \BibitemOpen
  \bibfield  {author} {\bibinfo {author} {\bibfnamefont {Y.}~\bibnamefont {Guo}}, \bibinfo {author} {\bibfnamefont {J.}~\bibnamefont {Pack}}, \bibinfo {author} {\bibfnamefont {J.}~\bibnamefont {Swann}}, \bibinfo {author} {\bibfnamefont {L.}~\bibnamefont {Holtzman}}, \bibinfo {author} {\bibfnamefont {M.}~\bibnamefont {Cothrine}}, \bibinfo {author} {\bibfnamefont {K.}~\bibnamefont {Watanabe}}, \bibinfo {author} {\bibfnamefont {T.}~\bibnamefont {Taniguchi}}, \bibinfo {author} {\bibfnamefont {D.}~\bibnamefont {Mandrus}}, \bibinfo {author} {\bibfnamefont {K.}~\bibnamefont {Barmak}}, \bibinfo {author} {\bibfnamefont {J.}~\bibnamefont {Hone}}, \bibinfo {author} {\bibfnamefont {A.~J.}\ \bibnamefont {Millis}}, \bibinfo {author} {\bibfnamefont {A.~N.}\ \bibnamefont {Pasupathy}}, \ and\ \bibinfo {author} {\bibfnamefont {C.~R.}\ \bibnamefont {Dean}},\ }\bibfield  {title} {\enquote {\bibinfo {title} {Superconductivity in twisted bilayer wse2},}\ }\href@noop {} {\ }\Eprint {http://arxiv.org/abs/arXiv:2406.03418}
  {arXiv:2406.03418} \BibitemShut {NoStop}%
\end{thebibliography}%

\end{document}